\documentclass[10pt,conference]{IEEEtran}

\usepackage{amsmath,amssymb,bbm,mathrsfs,bm,braket,color,graphicx,comment,xcolor} 
\usepackage[colorlinks,citecolor=blue,urlcolor=blue,linkcolor = blue]{hyperref}
\usepackage[mathscr]{euscript}
\usepackage{enumitem}
\usepackage{dsfont}
\usepackage{braket}
\usepackage{multirow}
\usepackage{float}
\usepackage{cite}
\usepackage{caption}

\begin{document}
\bstctlcite{IEEEexample:BSTcontrol}
\title{Constant Depth Code Deformations in the Parity Architecture}

\author{\IEEEauthorblockN{Anette Messinger}
\IEEEauthorblockA{\textit{Parity Quantum Computing GmbH}\\ A-6020 Innsbruck, Austria}
\and
\IEEEauthorblockN{Michael Fellner}
\IEEEauthorblockA{\textit{Parity Quantum Computing GmbH}\\ A-6020 Innsbruck, Austria\\\\
\textit{Institute for Theoretical Physics}\\ University of Innsbruck\\ A-6020 Innsbruck, Austria}
\and
\IEEEauthorblockN{Wolfgang Lechner}
\IEEEauthorblockA{\textit{Parity Quantum Computing GmbH}\\ A-6020 Innsbruck, Austria\\\\
\textit{Institute for Theoretical Physics}\\ University of Innsbruck\\ A-6020 Innsbruck, Austria}
}

\maketitle

\begin{abstract}
We present a protocol to encode and decode arbitrary quantum states in the parity architecture with constant circuit depth using measurements, local nearest-neighbor and single-qubit operations only. While this procedure typically requires a quadratic overhead of simultaneous qubit measurements, it allows for a simple and low-depth implementation of logical multi-qubit gates in the parity encoding via code deformation. We discuss how such encoding and decoding schemes can be used to flexibly change the size and shape of the underlying code to enable a more efficient implementation of quantum gates or algorithms. We apply the new findings to the QAOA which leads to a constant depth implementation using local gates at the same optimization performance as the standard, potentially non-local, QAOA approach without the parity encoding. Furthermore, we show that our method can reduce the depth of implementing the quantum Fourier transform by a factor of two when allowing measurements.
\end{abstract}

\section{Introduction}
The development of quantum computers has made huge steps in the last decades on the theoretical as well as on the experimental side, paving the way for demonstrating quantum advantage. The main achievements include quantum devices particularly designed for solving optimization problems~\cite{Johnson2011, Boixo2014, Nigg2017} as well as universal quantum computers~\cite{Chow2012, Saffman2016, Pichler2017, Postler2022}. However, there are still hardware limitations researchers have to tackle~\cite{Leymann2020, Fellous-Asiani2021, Stilck2021, Gonzales2022}. One the one hand, these include error-prone gates, which set a limit to the maximal number of gates feasible in a quantum circuit, while on the other hand quantum decoherence limits the maximal circuit depth. In the long term, these difficulties are expected to be tackled via error correction techniques~\cite{Shor1996, Fowler2012, Horsman2012, Terhal2015}. In the noisy intermediate-scale quantum (NISQ) era~\cite{Preskill2018}, however, quantum errors have to be mitigated or avoided by minimizing the number of gates and the circuit run time.

Recently, a universal gate set for the parity architecture~\cite{Lechner2015} was proposed, allowing for an efficient implementation of corner-stone quantum algorithms like the quantum Fourier transform~\cite{Fellner2022universal, Fellner2022applications}. In the parity encoding, operators diagonal in the $Z$-basis are implemented with physical single-qubit operations, while most non-diagonal operators require CNOT sequences of depth scaling linear with the system size. Similarly, the encoding and decoding processes presented in previous works require sequences of CNOT gates of linear depth.

\begin{figure*}
\includegraphics[width=\textwidth]{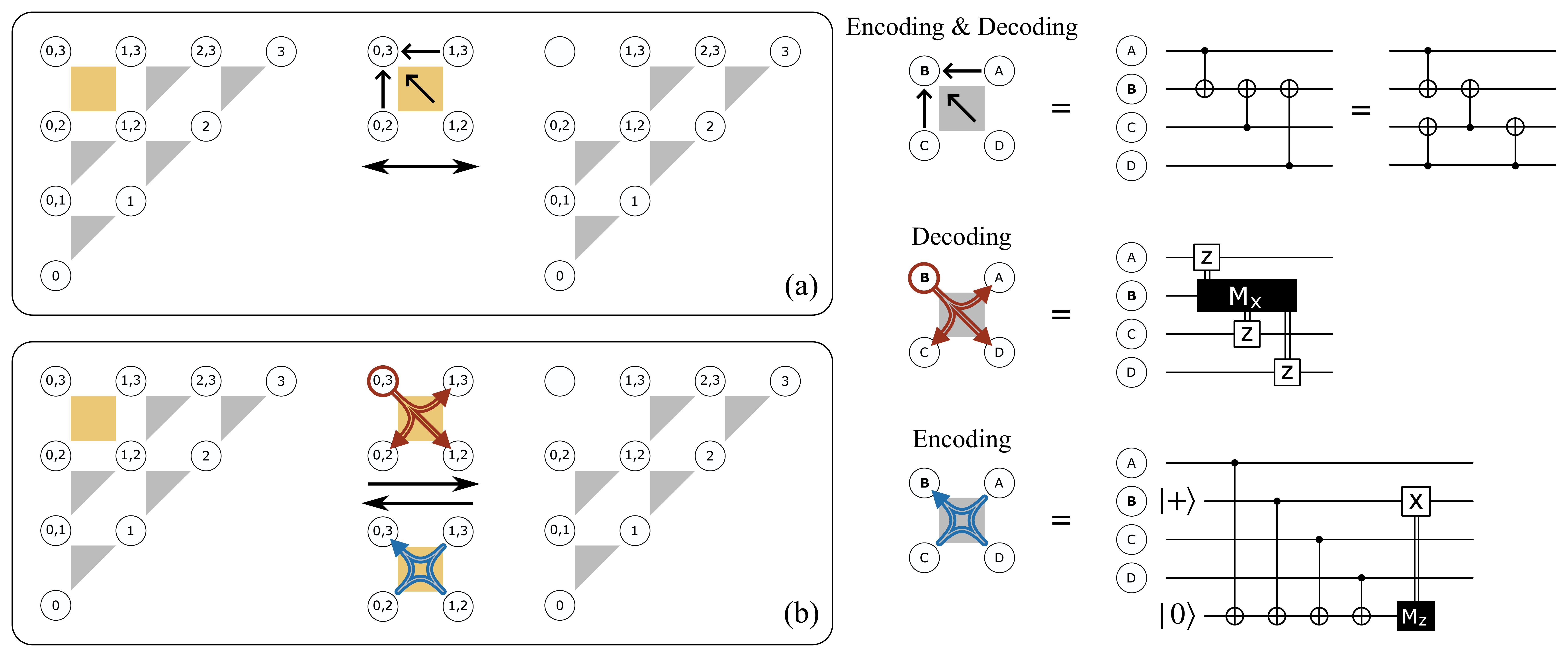}
\caption{Comparison of two different encoding and decoding schemes for a single parity qubit in the LHZ layout: (a) Using CNOT gates, the qubit indicated by an empty circle is in the state $\ket{0}$. (b) Using measurements and classical corrections. Here, the qubit indicated by an empty circle is in the state $\ket{+}$ before the encoding and in either $\ket{+}$ or $\ket{-}$ after decoding. The encoding requires an ancilla qubit initialized in the state $\ket{0}$ which could for example be placed in the center of the constraint (see Fig.~\ref{fig:layout}). For both cases, the constraint used for the de- or encoding is highlighted in yellow.}
\label{fig:scheme_comparisons}
\end{figure*}

In this work we introduce protocols related to methods known as lattice surgery and code deformation~\cite{Vuillot2019, Bombin2009} to replace these sequential CNOT gates by parallel measurements and single-qubit gates (see Fig. \ref{fig:scheme_comparisons}). In particular, we build upon the ideas of quantum state teleportation~\cite{Bennett1993teleportation} to encode states in the parity architecture in constant circuit depth by performing stabilizer measurements and applying bit-flips dependent on their outcome. Similarly, we show that decoding can be performed using single-qubit measurements and conditional phase-flip operations.
The constant depth implementation of the proposed encoding and decoding schemes implies that also non-diagonal gates can be implemented in the parity architecture with constant depth. For single-qubit gates this is achieved by decoding the system to act on, performing the gate on the logical qubits, and encoding again.

One of the most promising candidates to show quantum advantage in the NISQ-era in solving combinatorial optimization problems is the Quantum Approximate Optimization Algorithm (QAOA)~\cite{Farhi2014}.
As the parity mapping exhibits various benefits for the implementation for such optimization algorithms~\cite{Lechner2020, Fellner2021}, we apply our findings to the QAOA and in particular show how the new encoding approach leads to a constant-depth implementation of the QAOA in the parity architecture while implicitly preserving all parity constraints by using logical bit-flip operators in the mixing Hamiltonian~\cite{Hadfield2019}. So far, parity QAOA was either implemented in constant depth by enforcing (at least part of) the parity constraints via an additional energy penalty in the cost function~\cite{Lechner2020, Ender2022, Unger2022} or fulfilling the constraints implicitly at the cost of a linear circuit depth~\cite{Ender2022}. Our approach combines the advantages of the constant-depth parity QAOA and fully constraint preserving (implicit) parity QAOA by exhibiting the same Hilbert-space and classical search space as the optimization problem in the original formulation while featuring constant circuit depth.

The remainder of this paper is organized as follows. In Sec.~\ref{sec:parity}, we briefly review the basics of the parity code, on which we build our new approach. We then introduce the concept of code deformation in the parity architecture in Sec.~\ref{sec:encodingdecodingmeasurement}, and outline some applications in Sec.~\ref{sec:applications}. Finally, some conclusive remarks can be found in Sec.~\ref{sec:conclusion}.

\section{The Parity Code}\label{sec:parity}
The parity code~\cite{Lechner2015, Ender2021} can be described as a stabilizer code in which $n$ logical qubits are encoded into $K$ physical qubits such that single-body physical Pauli-$Z$ operations translate to products of logical $Z$ operations. Therefore, in the code space, computational basis states of physical qubits encode the \textit{parity} of the basis states of logical qubits. We label each physical qubit with the set $\mathcal{L}$ containing the indices $(i,j,\dots)$ of exactly these logical qubits, such that for any $\ket{\psi}$ in the code space\footnote{For better readability, we use round brackets for these labels whenever explicitly stating the logical indices. Note, however, that this does not imply any ordering.}
\begin{equation}\label{eq:parityrelation}
    Z_{(i,j,\dots)}\ket{\psi} = \tilde Z_i \tilde Z_j \dots \ket{\psi}
\end{equation}
holds. Here and in the following, we denote logical operators with a tilde. We further call physical qubits which encode the parity of at least two logical qubits \textit{parity qubits}, and physical qubits corresponding to only a single logical qubit \textit{data qubits}. While it is in principle not necessary to include data qubits in the code, their presence simplifies many operations and ensures that there is no ambiguity in the definitions of logical states. Unless otherwise stated, we therefore only consider code constructions which include a data qubit for every logical qubit in this work.

The stabilizer of this code is generated by a set of parity constraints of the form $\prod_i Z_{\mathcal{L}_i}$,
where $\mathcal{L}_i$ are the sets of logical qubit indices labelling physical qubits such that every logical index appears an even number of times in the whole product, i.e., the symmetric difference of all sets in the constraint  is the empty set,
\begin{equation}
    \mathcal{L}_1 \ominus \mathcal{L}_2 \ominus \dots = \emptyset.
\end{equation}
Equation~\eqref{eq:parityrelation} implies that a valid code state is an eigenstate of all constraints with eigenvalue $+1$ by construction,
\begin{equation}
    \prod_i Z_{\mathcal{L}_i}\ket{\psi} = \prod_i\prod_{j \in \mathcal{L}_i} \tilde Z_j\ket{\psi}=\ket{\psi},
\end{equation}
as all duplicate logical operators cancel each other. The smallest set of generators must contain ${K-n}$ independent constraints (independent in the sense that no constraint can be represented by the product of any others). For many logical circuits, there exists a layout for the physical qubits along a two-dimensional square lattice such that the stabilizer can be generated by a set of geometrically local constraints. As an example, the illustrations on the left side of Fig.~\ref{fig:scheme_comparisons}a-b show the so-called LHZ layout, which encodes a parity qubit for every two-body parity of logical qubits and additionally a data qubit for every logical qubit. In this layout, every parity constraint occupies only a ${2\times 2}$ square or triangle (plaquette) of qubits on the layout.

\subsection{Logical operations}
In the following we define logical operators in the physical Hilbert space. According to Eq.~\eqref{eq:parityrelation}, we define the logical Pauli-Z operator as
\begin{equation}
    \tilde Z_i = Z_{(i)}.
\end{equation}
Note that this operation is directly implementable if and only if the corresponding data qubit is part of the code. Similarly, arbitrary collective Z-rotations of multiple logical qubits 
\begin{equation}
\tilde R_{Z_i Z_j\dots}(\alpha)=\exp\left({-i\frac{\alpha}{2} \tilde Z_i \tilde Z_j \dots}\right)
\end{equation}
can be implemented due to their equivalence to the physical single-qubit rotation (up to multiplication with stabilizer generators) as
\begin{equation}
R_{Z_{(i, j, ...)}}(\alpha)=\exp\left({-i\frac{\alpha}{2} Z_{(i,j,\dots)}}\right)
\end{equation}
whenever the corresponding parity qubit is part of the code.

The logical Pauli-X operator, which preserves the code-space and fulfils the necessary commutation relations can be defined as
\begin{equation}\label{eq:logicalX}
    \tilde X_i = \prod_{j\in Q_i} X_{(j)},
\end{equation}
where $Q_i$ denotes the set of all physical qubits involving the logical index $i$.
In other words, a logical bit-flip is realized by simply flipping all physical qubits whose labels include the corresponding logical index. 

Arbitrary logical rotations about the X-axis,
\begin{equation}
    \tilde R_{X_i}(\alpha) = \exp\left(-i\frac{\alpha}{2} \tilde X_i\right), 
\end{equation}
can be implemented with physical CNOT gates along a tree which connects all involved physical qubits. See Ref.~\cite{Fellner2022universal} for a full description of a universal logical gate set.

\subsection{Encoding and Decoding}\label{subsec:CNOTencoding}
The easiest way to create a valid code state of a given set of parity and data qubits is to initialize all physical qubits in the state $\ket{0}$. This state is stabilized by all parity constraints, and corresponds to all logical qubits in the state $\ket{0}$. However, it is also possible to encode an arbitrary logical state from a physical state by defining the qubits describing that state as the data qubits of the code. As there are no constraints required within a set of exclusively data qubits, this is already a valid code state, although it is just the trivial code.
We can then add parity qubits one-by-one by encoding them in the state corresponding to the desired parity. Every added parity qubit must always be accompanied by a constraint which fixes the parity of the new qubit with respect to other qubits of the code\footnote{This only applies if the corresponding parity can be represented in terms of the existing qubits. If the parity qubit relates to a logical qubit which has not appeared in the code yet, the size of the logical Hilbert space is also increased and thus, no constraint is necessary.}. Before focusing on the measurement-based encoding and decoding strategy in the remainder of this paper, we briefly introduce the method presented in Ref.~\cite{Fellner2022universal}, which makes use of CNOT gates between physical qubits of the code and the newly added or removed qubit to encode or decode parity information.  Both the CNOT-based and the measurement-based method are illustrated in Fig.~\ref{fig:scheme_comparisons}.

As decoding is achieved by the same gate sequence as encoding, we focus only on encoding in the remainder of this section. The new qubit must first be initialized in the state $\ket{0}$. Then, after identifying a constraint which relates the new parity qubit to the existing code qubits, CNOT gates are applied from each of the other qubits in that constraint (as control) to the new qubit (as target), as depicted in Fig.~\ref{fig:scheme_comparisons}a. This method is also commonly applied for other stabilizer codes: Any new qubit with a stabilizer operator consisting solely of Z-terms can be added or removed with such a sequence of CNOT gates. Note that multiple different choices for such a constraint might exist, which gives the user the freedom to choose whichever constraint results in the least effort to implement the CNOT gates. Ref.~\cite{Fellner2022applications} suggests a number of applications for which an efficient placement of physical qubits exists such that there always are local constraints implementable by nearest-neighbor interactions without additional overhead. For the LHZ layout considered in Fig.~\ref{fig:scheme_comparisons}, for example, the parity qubits can be added in layers (diagonal in the figure) from the first layer of data qubits. Every new parity qubit can then be added to the code with the help of a three- or four-body constraint of qubits in an adjacent square unit cell connecting the new qubit to the previous layer, as described in the supplemental material of Ref.~\cite{Fellner2022universal}. The drawback of this method is that the circuit depth for such an implementation scheme typically grows with the problem size. This is due to the fact that a local constraint fixing a new parity qubit can depend on other parity qubits being encoded already and thus not all the CNOT gates for the encoding can be applied in parallel.

Depending on the layout of qubits which are encoded, some improvements to the gate count and circuit depth are possible. For example, in some cases the same CNOT gate can be used to help encode multiple parity qubits at once. As there is a lot of freedom in the choice of the CNOT sequence (see for example the two alternative circuits representing the scheme of Fig.~\ref{fig:scheme_comparisons}a), such situations can be favoured with a clever choice of CNOT gates. In order to directly construct a more efficient encoding circuit, it helps to look the the action of a CNOT gate at the level of computational basis states, 
\begin{equation}
    \text{CNOT}:\hspace{3mm} \ket{a}_c\ket{b}_t \mapsto \ket{a}_c\ket{a\oplus b}_t,
\end{equation}
where $c$ and $t$ denote the control and target qubit, respectively, ${a, b\in \{0, 1\}}$, and $\oplus$ denotes the addition modulo 2.
Parity qubits can be described similarly in dependence of their corresponding logical qubits. If a set of logical qubits is in the states $\ket{a_i}, \, \ket{a_j},\dots$ the state of the corresponding parity qubit must be $\ket{a_{(i,j,\dots)}}$ with
\begin{equation}
    a_{(i,j,\dots)} = a_i\oplus a_j\oplus \dots .
\end{equation}
Therefore, a valid encoding procedure can always be constructed by applying a sequence of CNOT gates that collects the desired logical information from the data qubits and encodes it in the respective parity qubits.

\section{Encoding and Decoding with measurements}\label{sec:encodingdecodingmeasurement}

In this section, we introduce an alternative way to change the code, which is beneficial if a large chunk of qubits on the layout is to be en-/decoded at once. This new approach uses measurements and classical corrections depending on the measurement outcomes, and can be understood as a form of code deformation~\cite{Vuillot2019} as has also been applied for example in the surface code~\cite{Fowler2012}. 

\begin{figure}
    \centering
    \includegraphics[width=.95\columnwidth]{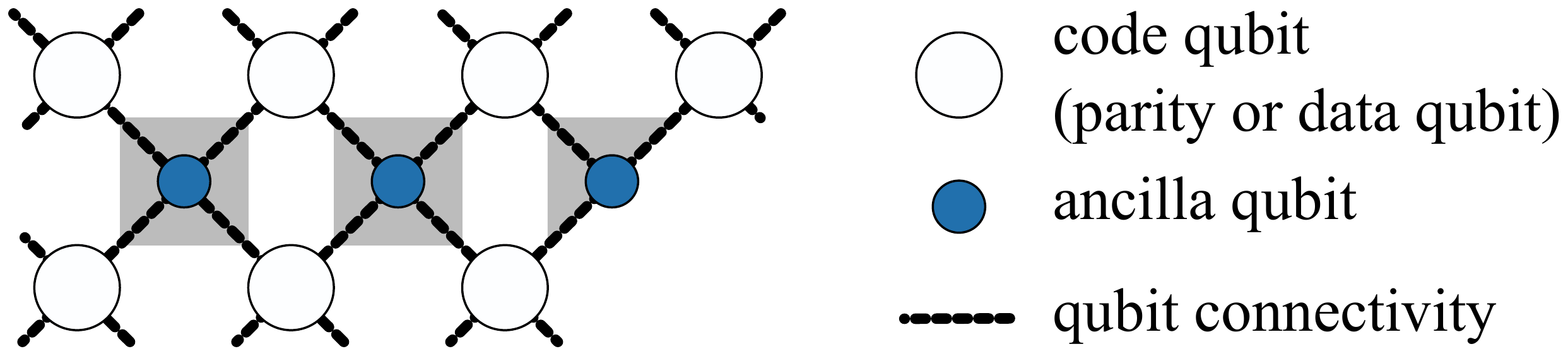}
    \caption{Possible placement of ancilla qubits (blue) for constraint measurements. The minimal required qubit connectivity is still along a square-lattice (but rotated). If physical interactions between neighboring code qubits are required but not available, they can also be mediated by an ancilla qubit.}
    \label{fig:layout}
\end{figure}

For the encoding process, we start with a new qubit in the state $|+\rangle$. We then choose a constraint which fixes the parity qubit we want to encode relative to already existing parity or data qubits and measure the value of this constraint. If the constraint measurement indicates a violation of the constraint, we perform a bit-flip operation on the new qubit. The constraint measurement can, for example, be performed on an ancilla qubit following a sequence of CNOT gates targeting the ancilla and controlling each of the constraint qubits (see Fig.~\ref{fig:scheme_comparisons}, bottom circuit). For encodings like the LHZ layout, where each constraint occupies at most a 2 x 2 cell of qubits (see also Ref.~\cite{Ender2021} for more general layouts of this form), the ancillas for constraint measurement can be placed in the center of that cell, as indicated in Fig.~\ref{fig:layout}.

For the decoding process, we directly measure the qubit we want to remove in the $X$ basis, and apply a conditional phase-flip operation on all other qubits of a constraint which fixed the removed qubit (see Fig.~\ref{fig:scheme_comparisons}, middle circuit). Note that this can be the same constraint which was used to encode the qubit, but it can in principle be any set of parity or data qubits with the same parity as the removed qubit. As the communication in this protocol is  purely classical and non-local classical operations are not a problem, the most straightforward choice is the constraint between the removed qubit and the corresponding set of data qubits.

\subsection*{Equivalence of protocols}
\begin{figure}
    \centering
    \includegraphics[width=\columnwidth]{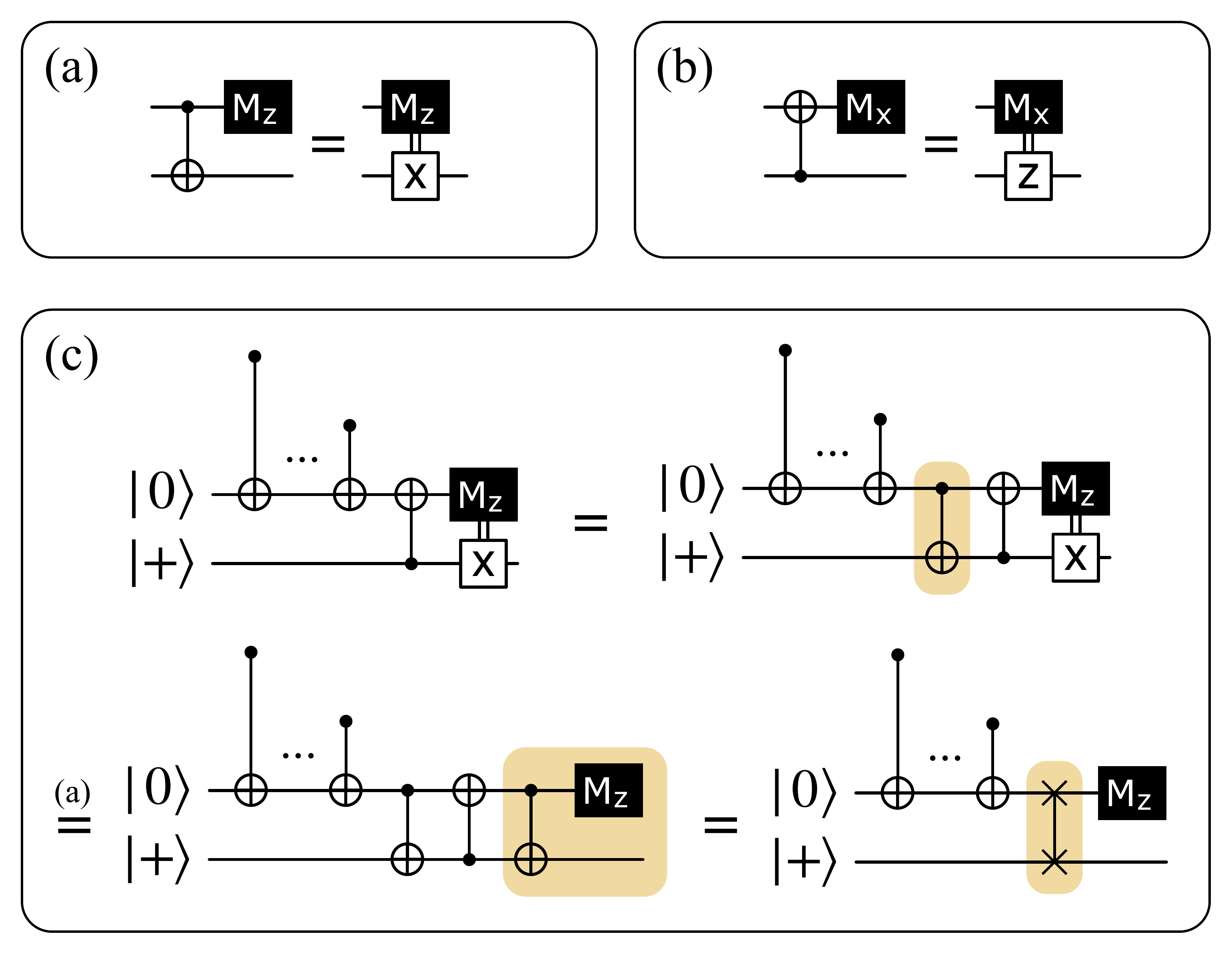}
    \caption{Circuit identities to derive measurement-based encoding and decoding schemes. (a-b) Classical versions of the CNOT gate in the $Z$- and $X$-basis, which are equivalent to the corresponding quantum gate if followed by the corresponding measurement. The decoding scheme can be derived by adding a measurement operation on the decoded qubit (which has no effect on the state of the remaining code qubits) and using identity (b) for each CNOT gate of the original decoding circuit. (c) Derivation of the encoding scheme with an ancilla qubit initially in the state $\ket 0$. In the first step, a CNOT gate can be added without effect because the target qubit is in the state $\ket{+}$, the second step uses the identity (a) and in the last step the CNOT sequence is interpreted as a SWAP gate.}
    \label{fig:teleportation}
\end{figure}

Figure~\ref{fig:teleportation} shows a derivation of the equivalence of the measurement-based encoding and decoding schemes to the gate-based scheme proposed in Ref.~\cite{Fellner2022universal}. The decoding process can be understood as the classical analog of the corresponding CNOT sequence\footnote{For a direct analogy, consider the phase-kickback action of the CNOT gate in the $X$-eigenbasis: It performs a phase flip on the \textbf{control} qubit if the \textbf{target} qubit is in the state $\ket{-}$.}, as introduced in Sec.~\ref{subsec:CNOTencoding}. If the control qubit of a CNOT gate is measured after executing the gate, the CNOT gate can be replaced by a conditional bit-flip operator on the target qubit which is classically controlled by the measurement outcome (see Fig.~\ref{fig:teleportation}a). As decoded qubits are disentangled from the code qubits and thus not required anymore, we can measure these qubits without information loss and use the same idea of replacing the CNOT gate by a classically controlled operator, but now in the $X$-basis (i.e., measuring the target qubit, see Fig.~\ref{fig:teleportation}b), to replace the CNOT gates of the original decoding algorithm by classically-controlled phase-flip operations.

The measurement-based encoding scheme can be understood by considering the concept of state teleportation~\cite{Bennett1993teleportation}: Instead of directly encoding the new parity qubit we first encode the new parity onto an ancilla qubit and then teleport its state onto the new parity qubit (see Fig.~\ref{fig:teleportation}c for an illustrative derivation of the state teleportation effect).

\subsection*{Simultaneous measurements and classical corrections}
The main advantage of this measurement-based strategy is that an arbitrary number of parity qubits can be simultaneously encoded or decoded in constant depth, irrespective of  dependencies between the parity qubits (via the constraints used for the encoding or decoding). 
While for the CNOT-based method, one has to add or remove the qubits one-by-one in the correct order (such that every added or removed qubit can be fixed using currently existing parity qubits in the geometrical vicinity), in the measurement-based method we can apply all the gates and measurements in parallel. The conditional spin- or phase flips then depend on multiple measurement outcomes, but the processing of this information is purely classical. Note that every constraint can only be used to encode or decode a single qubit.
It is, however, allowed to choose a constraint which also contains other qubits which are added or removed in the same step (using other constraints).

\begin{figure}
    \centering
\includegraphics[width=\columnwidth]{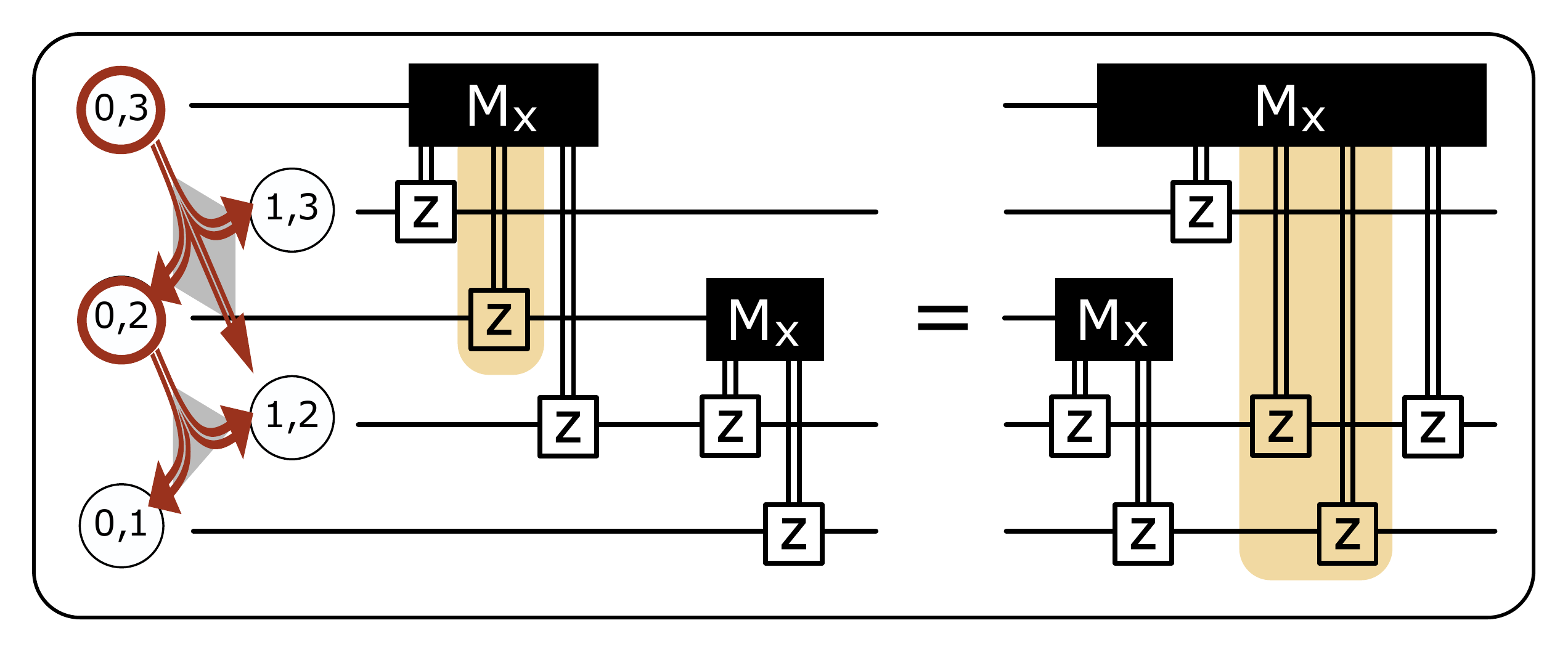}
    \caption{Commutation relation of an $X$-measurement and the corrective phase flip induced by another $X$-measurement when decoding the qubits ${(0,3)}$ and ${(0,2)}$. The affected corrective phase flip and the resulting flips after commuting are highlighted in yellow. Note that in a further optimization step, the corrections on qubit ${(1,2)}$ due to the measurement on qubit ${(0,3)}$ can be cancelled.}
    \label{fig:measurement_commutation}
\end{figure}

In a sequential implementation of the measurement-based encoding scheme, the outcome of every constraint measurement which contains more than one new qubit also depends on whether some of these new qubits were flipped by their respective encoding constraint measurements. As in the constant-depth implementation this correction takes place after all the constraint measurements, we need to reinterpret the outcome of every measurement taking into account all corrective flips which could affect the corresponding constraint. For example, consider removing qubits ${(0,2)}$ and ${(0,3)}$ from the LHZ layout shown in Fig.~\ref{fig:scheme_comparisons} at the same time using constraints ${\bm{(0,3)}-(0,2)-(1,3)-(1,2)}$ and ${\bm{(0,2)}-(1,2)-(0,1)}$ with simultaneous measurements (see also Fig.~\ref{fig:measurement_commutation}). As there is a corrective $Z$-flip on qubit ${(0,2)}$ conditioned  on the measurement outcome on qubit ${(0,3)}$, the $Z$ flip condition for qubits ${(0,1)}$ and ${(1,2)}$ [which is the outcome on ${(0,2)}$] must be replaced by the product of the outcomes on ${(0,2)}$ and ${(0,3)}$. Mathematically, this corresponds to commuting the corrective flips with the measurement, as is shown in Fig.~\ref{fig:measurement_commutation} for the above example.

\begin{figure}
    \centering
\includegraphics[width=\columnwidth]{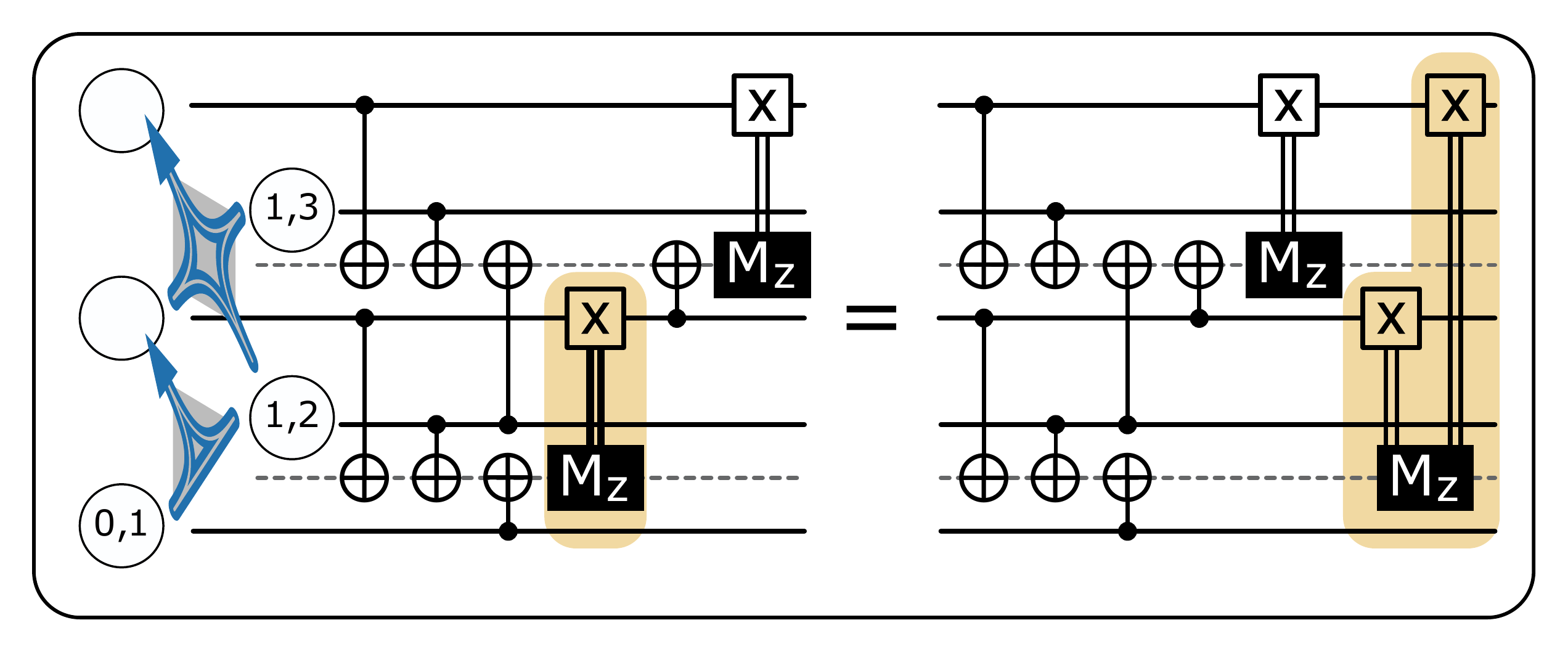}
    \caption{Commutation relation of a $Z$-measurement and the corrective bit flip induced by another $Z$-measurement when encoding the qubits ${(0,3)}$ and ${(0,2)}$. Dashed lines correspond to ancilla qubits used for constraint measurements. The affected corrective bit flip and the resulting flips after commuting are highlighted in yellow.}
    \label{fig:enc_measurement_commutation}
\end{figure}

Similarly, when removing the two qubits from the code, a sequential implementation would require qubit ${(0,2)}$ to be added before qubit  ${(0,3)}$. Full parallelization would be blocked by the corrective bit-flip on qubit ${(0,2)}$. Even though this qubit is not measured for adding the next qubit, it still influences the corresponding constraint measurement via a CNOT gate. This influence must be taken into account in the fully parallel implementation, where all measurements are performed before the corrections (see Fig.~\ref{fig:enc_measurement_commutation}).

The final corrections for arbitrarily large code manipulations can always be calculated iteratively (and classically) from the full set of measurement outcomes.
For this, we define the set $S$ as the set of qubits for which we have determined the necessary corrections.

Initially, $S$ contains all qubits which are not subject to any corrections.
In the encoding process, these are precisely the qubits which were already in the code before the encoding. In the decoding process, this includes the qubits to be decoded on which no other qubits depend via their respective decoding constraints. Furthermore, it can include qubits which remain in the code and are not in any of the decoding constraints.
We then apply the following steps until $S$ contains all qubits:
\begin{itemize}
    \item[1.] Determine the corrections to the set $S'$ of all qubits whose correction depends only on measurement results of qubits in $S$, taking into account their determined corrections.
    \item[2.] Update $S$ by reassigning $S \gets S\cup S'$.
\end{itemize}

\begin{figure*}
    \centering
    \includegraphics[width=0.3\textwidth]{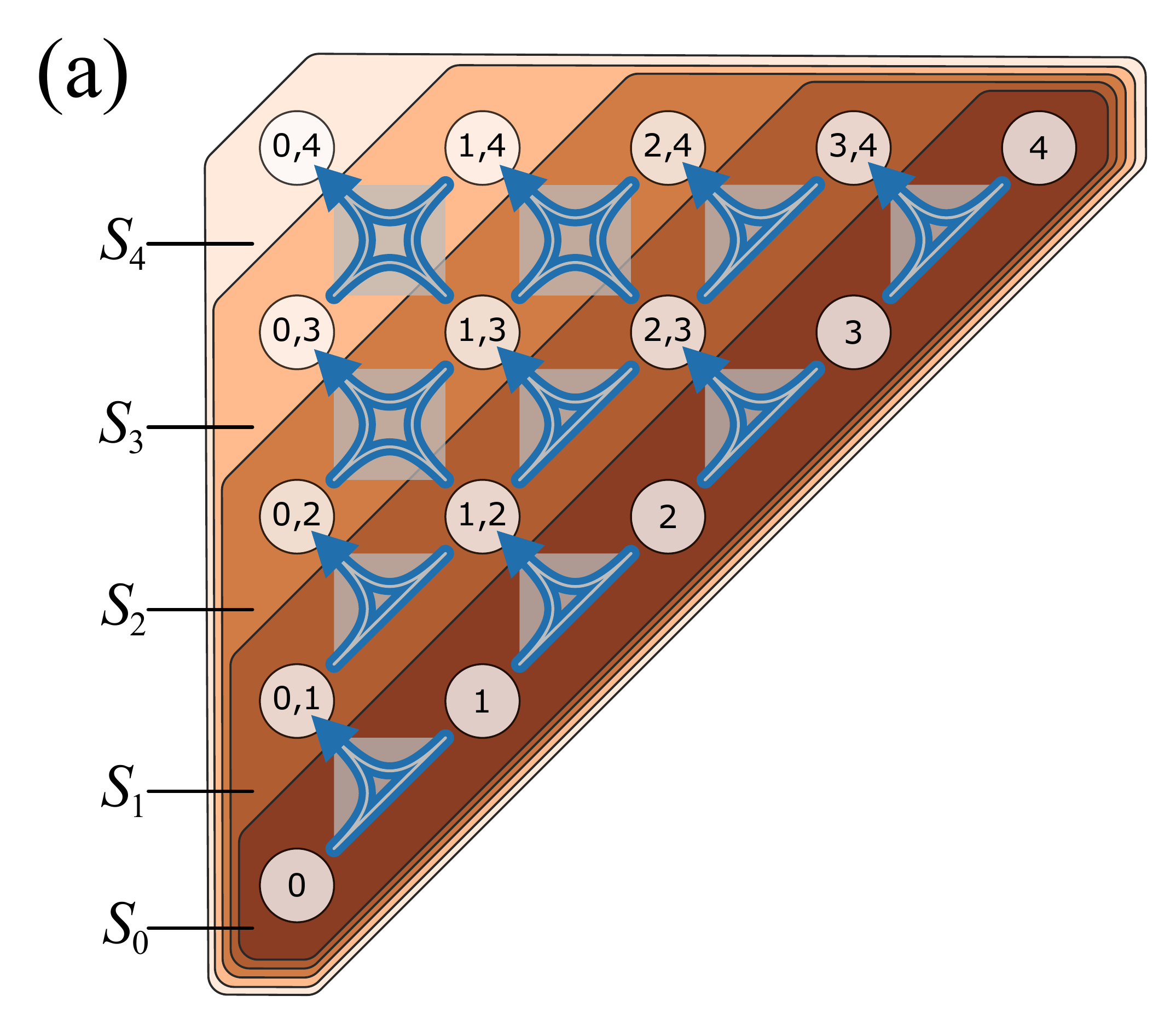}
    \hspace{.5cm}
    \includegraphics[width=0.3\textwidth]{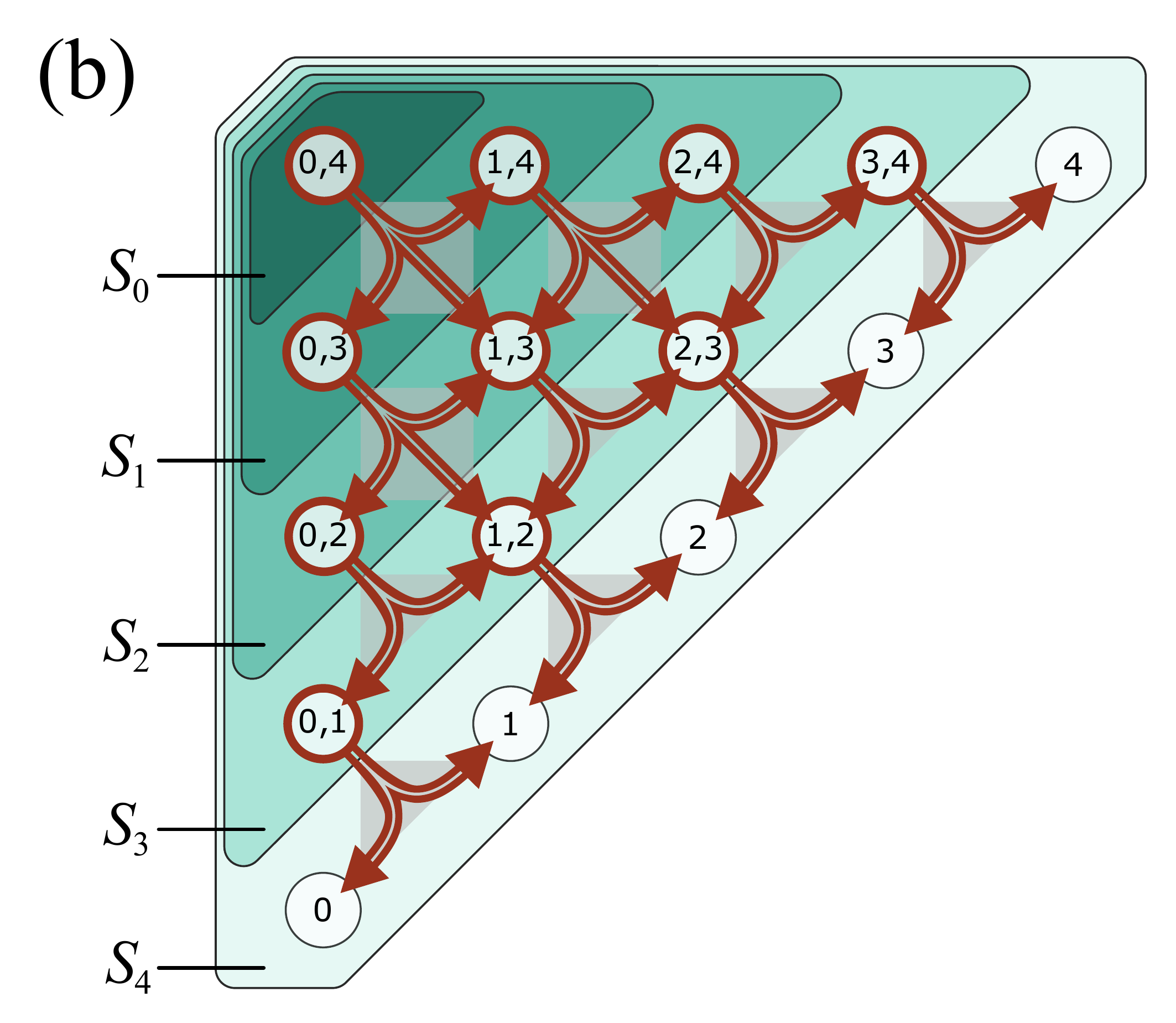}
    \hspace{.5cm}
    \includegraphics[width=0.3\textwidth]{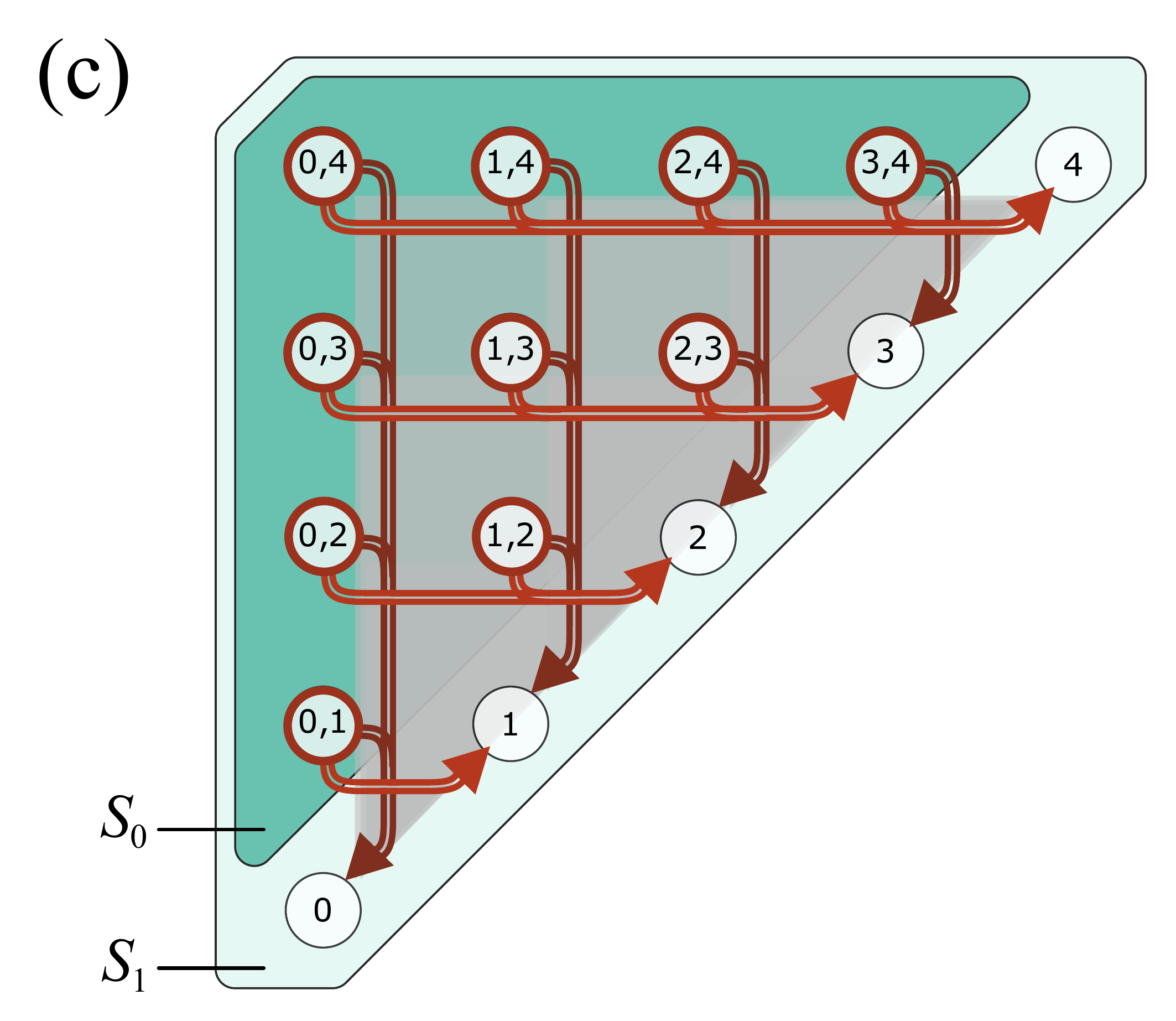}
    \caption{Classical information flow to determine qubit corrections for encoding (a) or decoding (b) of a fully connected logical system using simultaneous measurements. The colored regions contain the qubits in the set $S$ at each time step of the classical correction process, ${S_0\subset S_1 \subset S_2 \subset S_3 \subset S_4}$. Panels (b) and (c) show two different choices of constraints used for decoding. In panel (c), all parity qubits are decoded directly with respect to their corresponding data qubits, such that all classical corrections can be calculated in a single iteration.}
    \label{fig:encoding_decoding_corr}
\end{figure*}

For example, when encoding the complete triangular LHZ layout from only the data qubits, every step adds one diagonal row to $S$, starting at the row of data qubits and finishing at the parity qubit at the opposite tip of the triangle (see Fig.~\ref{fig:encoding_decoding_corr}a). That is, we first determine the corrections to qubits ${(0,1)}, {(1,2)}, {(2,3)},\dots$, then the corrections to qubits ${(0,2)}, {(1,3)}, \dots$ and so on until qubit ${(0,N-1)}$.
Reversely, in the decoding process, we start with only the parity qubit at the tip of the triangle, ${(0,N-1)}$, and in each step add a row towards the final row of data qubits (see Fig.~\ref{fig:encoding_decoding_corr}b). Alternatively, one can choose different constraints for the decoding and directly correct the corresponding data qubits for every decoded parity qubit, as shown in Fig.~\ref{fig:encoding_decoding_corr}c.

Note that in any case, the total depth of the quantum circuit is only that of a single constraint or qubit measurement and a single-qubit gate. All other logic is executed classically.

For the LHZ-layout and constraints as depicted in Fig.~\ref{fig:encoding_decoding_corr}, the exact formulas for the corrections after measurements can be derived as follows. Let ${m_{ij}^z\in \{\pm 1\}}$ denote the result of the $Z$-measurement of the constraint used to encode the qubit labelled ${(i, j)}$, with ${i, j \in\{0, 1, \dots, n-1\}}$.  We further define the spin variable
\begin{equation}
    s_{kl}:=\prod_{i=k}^{l-1}\prod_{j=i+1}^l m_{ij}^z.
\end{equation}
corresponding to the product of all measurement results that affect parity qubit $(k,l)$. Then the correction that must be applied to qubit $(k,l)$ is

\begin{equation}
    \mathcal{C}_\mathrm{enc}^{kl}= \left[X_{(k,l)}\right]^{\frac{1}{2}(1-s_{kl})}.
\end{equation}

For full decoding, let $m_{ij}^x$ denote the result of an $X$-measurement of parity qubit ${(i,j)}$. The correction operator to be applied to data qubit $(i)$ is then
\begin{equation}
    \mathcal{C}_\mathrm{dec}^{i}=\left[Z_{(i)}\right]^{\frac{1}{2}(1-s_{i})},
\end{equation}
with 
\begin{equation}
    s_i = \prod_{l=0}^{i-1} m_{li}^x\prod_{j=i+1}^{n-1} m_{ij}^x.
\end{equation}

\subsection{Higher order parities}
\begin{figure}
    \centering
    \includegraphics[width=0.5\columnwidth]{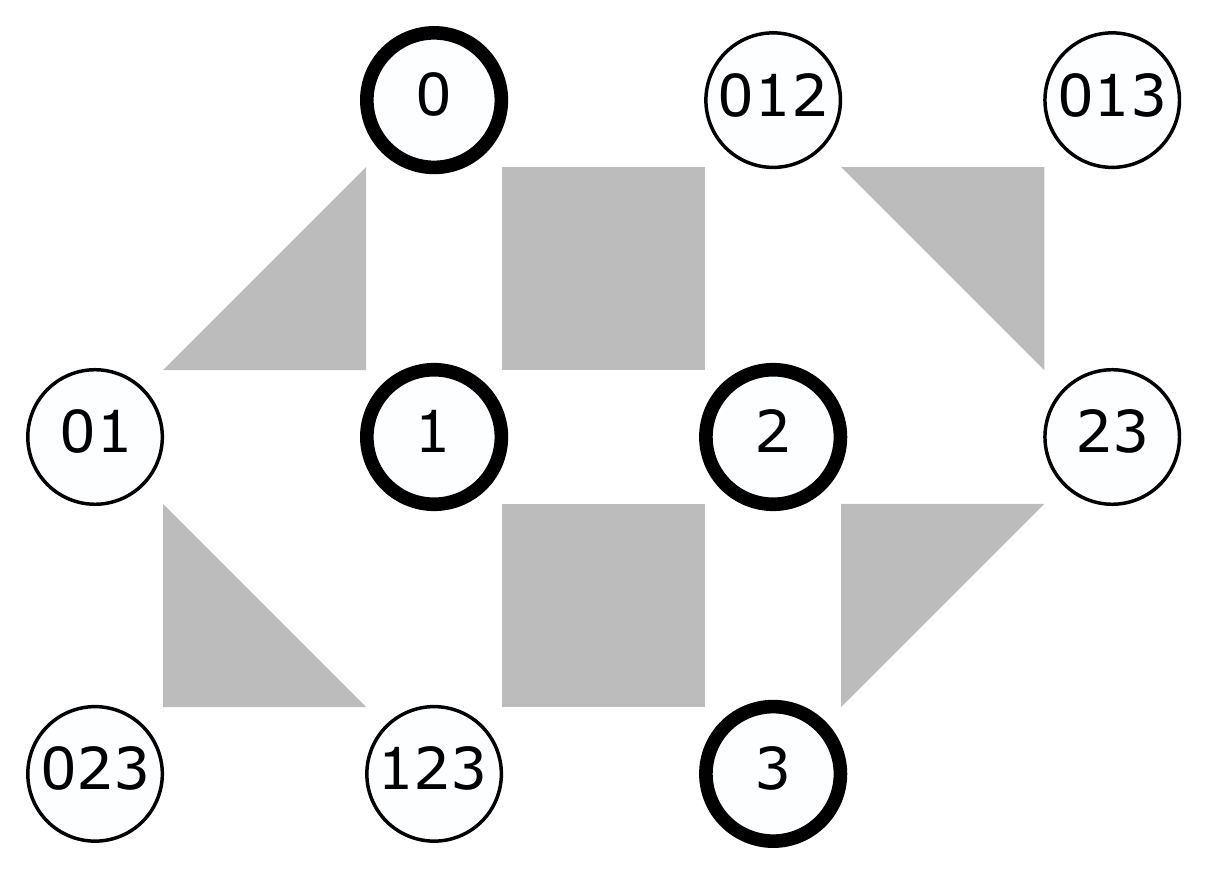}
    \caption{Example of constraint layout to encode three-body interactions. The parity qubits ${(2,3)}$ and ${(0,1)}$ are ancillary and only required 
    to obtain an encoding of all three-body terms using local constraints. With the data qubits (bold) as a basis, all parity qubits can be added or removed from the code in constant depth.}
    \label{fig:three-body}
\end{figure}

The encoding and decoding protocol can be formulated completely in terms of the stabilizers of the code, independent of the exact mapping to the logical qubits. This includes mappings in which some qubits hold the parity of more than two qubits. Encoding higher-order parity qubits is not always straightforward, but it has been shown~\cite{terhoeven2023} that, given enough ancillary qubits, one can always find a valid layout with local constraints on four or less qubits which includes the desired parity qubits (see Fig.~\ref{fig:three-body} for an example for all possible three-body parities on four qubits). Therefore, it is also possible to use the same constant-depth circuits for code layouts with arbitrarily high-order parity. Note that, while the quantum circuit still has constant depth, the classical complexity to calculate the corrections can then scale polynomial with the number of logical qubits.

\subsection{Partial Encoding and Decoding}
Instead of encoding the full set of parity qubits in the beginning of the algorithm and decoding all of them in the end, one can also use the encoding and decoding strategies more flexibly to adjust the set of parity qubits to the algorithm on-the-fly by decoding only some of them and potentially encoding different ones whenever needed. One particularly interesting scenario is the removal of all parity qubits contained in a logical Pauli-X operator [see Eq.~\eqref{eq:logicalX}] in order to directly perform non-diagonal operations on the corresponding logical qubit. 

In general, any set of physical qubits can be removed from the code as long as the qubits which remain in the code form a valid readout basis spanning the full logical space, i.e., it is possible to deduct the state of every logical qubit from the remaining physical qubits.
It is in principle even possible to remove data qubits from the code, while preserving the corresponding logical qubit information indirectly through other parity qubits. 
For example, qubits ${(0)}$ and ${(0,1)}$ are sufficient to determine the state of logical qubit 1, whose computational basis states are described by the parity of the two physical qubits. 
Whenever there is a constraint with which a qubit can be decoded, it is valid to do so, even if the decoded qubit was a data qubit. This can be easily verified as one can use the same constraint to encode the qubit again, so no logical information is lost.

\subsection{Implications for Fault Tolerance}
Every decoding process produces a code with potentially less fault tolerance, as the code distance (typically the smallest number of physical qubits which share a logical index) can decrease when reducing the number of physical qubits. Note that in many cases, the code distance can remain the same, for example when removing a parity qubit which does not include any of the logical indices which determine the distance (i.e., which have the smallest number of occurrences in parity qubits).
In addition, replacing the CNOT-based protocols with measurements and classical corrections can lead to higher error rates due to faulty measurements, depending on the physical properties of the chosen platform. During the encoding process, such measurement errors can be mitigated by performing each constraint measurement several times and using a classical error syndrome decoder to identify measurement errors and determine the correct action.

\section{Applications}\label{sec:applications}
In the following we present a selection of applications of universal parity quantum computing for which the measurement-based approach can be used to improve the circuit depth.

\subsection{Measurement-based parity QAOA}
The Quantum Approximate Optimization Algorithm (QAOA)~\cite{Farhi2014} is a hybrid quantum-classical algorithm to solve combinatorial optimization problems.
The core idea of the QAOA is to prepare a solution candidate state
\begin{equation}
    \ket{\psi(\bm{\beta}, \bm{\gamma})} = \prod_{j=1}^p e^{-i\beta_j H_X} e^{-i\gamma_j H_P}\ket{+}^{\otimes n},
\end{equation}
where $H_X$, diagonal in the $X$-basis, represents a so-called mixer Hamiltonian and $H_P$, diagonal in the $Z$-basis, corresponds to the problem Hamiltonian of the optimization problem treated. 
The generic forms of these Hamiltonians are
\begin{equation}
H_X= \sum_i X_i
\end{equation}
and 
\begin{equation}
H_P= \sum_i J_i  Z_i + \sum_{i<j} J_{ij} Z_i Z_j+\sum_{i<j<k} J_{ijk}  Z_i Z_j Z_k+\dots,
\end{equation}
respectively. The integer $p$ refers to the number of QAOA layers and the $2p$ classical parameters ${\bm{\beta}=(\beta_1, \dots, \beta_p)}$ and ${\bm{\gamma}=(\gamma_1, \dots, \gamma_p)}$ are optimized in a quantum-classical feedback loop in order to minimize the energy expectation value ${\braket{H_P}=\braket{\psi|H_P|\psi}}$.
In standard approaches to implement the QAOA, the number, order and non-locality of interactions pose a limit to the parallelizability of the corresponding quantum circuit as many operations need to be decomposed into native gates of the underlying architectures. Also a mapping to quadratic unconstrained binary optimization (QUBO) problems~\cite{Lewis2017} cannot overcome these issues. The parity architecture, however, can be used to map all many-body interactions of the problem Hamiltonian $H_P$ onto local operations. 

In earlier works~\cite{Lechner2020}, only the problem Hamiltonian was mapped to the parity code and implemented as ${\tilde H_P= \sum_{k} J_{\mathcal{L}_k} Z_{\mathcal{L}_k}}$, while the mixer Hamiltonian was replaced by a mixer Hamiltonian with single-qubit $X$-operators on all physical qubits. In such an implementation, satisfaction of the parity constraints is not guaranteed and thus has to be enforced via an additional energy penalty. This increase of dimension of the reachable Hilbert space is avoided by also mapping the mixing Hamiltonian such that the whole evolution is restricted to the code space~\cite{Hadfield2019, Rocchetto2016}. However, the resulting mixing operators can become expensive to implement~\cite{Ender2022, Fellner2022applications} within the parity code, leading to a circuit depth scaling linearly with the system size.

In the following, we apply the measurement-based encoding and decoding techniques to perform the mixing operators on the logical qubits in the decoded state, using $\tilde H_X=\sum_k X_{(k)}$, while the parity-mapped problem Hamiltonian is implemented on the physical qubits in the encoded state.
We show how this leads to a constant QAOA circuit depth without the need for parity constraints in the problem Hamiltonian. 
Note that, if all data qubits are present in the parity code, this corresponds to a fully mapped implementation of the original problem and thus the energy landscape of $\braket{H_P}$ as a function of $\bm{\beta}$ and $\bm{\gamma}$ is equivalent to the energy landscape of the logical problem and therefore preserves useful properties like, for example, parameter concentration~\cite{Brandao2018, Streif2020, Akshay2021}.

\subsubsection{LHZ-scheme with data qubits}
In the $n$-qubit LHZ-scheme providing all two-qubit interactions as well as the corresponding data qubits (i.e.,\ ${K=n(n+1)/2}$ physical qubits in total), the procedure to perform QAOA using the measurement-based encoding approach is straight-forward and executed by the following procedure:

\begin{enumerate}
    \item Prepare the data qubits in the state $\ket{+}^{\otimes n}$.
    
    \item \textbf{for} ${1\leq j \leq p}$ \textbf{do}
    \begin{enumerate}
        \item Perform the measurement-based encoding sequence to build up the whole LHZ triangle.
        \item Apply the mapped problem Hamiltonian unitary
        \begin{equation}
            \tilde U_P=e^{-i\gamma_j \tilde H_P}=\prod_{k=1}^K \exp\left(-i\gamma_j J_{\mathcal{L}_k} Z_{\mathcal{L}_k}\right)
        \end{equation}
        on the corresponding physical qubits.
        \item Perform the measurement-based decoding sequence to collapse all quantum information to the data qubits, which after that represent the logical qubits.
        \item Apply the  mixing unitary
        \begin{equation}
            \tilde U_X=e^{-i\beta_j \tilde H_X}=\prod_{k=1}^n \exp\left(-i\beta_j X_{(k)}\right)
        \end{equation}
        on the data qubits.
        \item Reassign ${j \gets j+1}$.
    \end{enumerate}
    \item Evaluate the energy expectation value $\braket{H_P}$ and update the classical parameters.
    \item If a stopping criterion is reached, terminate. Otherwise go to step 1.
\end{enumerate}

\subsubsection{General parity layouts}
The same approach can be applied to any other parity code as long as there is a data qubit for every logical qubit.
For more general optimization problems, not all single-body terms must be present in the problem Hamiltonian and thus not all data qubits are necessarily part of the encoding. In that case, another valid readout basis ${\{\mathcal{L}_i| \, i=1, \dots,n\}}$ of the existing parity and data qubits can be chosen which defines the logical space. As the choice of the driver Hamiltonian in QAOA has a certain degree of freedom, it is not necessary to realize exactly the logical Hamiltonian ${\tilde H_X=\sum_{i=1}^n\tilde X_i}$. 
Instead, decoding to the chosen readout basis and then applying the Hamiltonian $ {\tilde H_X^\prime=\sum_{i=1}^n X_{\mathcal{L}_i}}$ acting on all physical qubits of the readout basis is also a valid choice. In the logical system, this corresponds to replacing at least some of the operators ${\tilde X_i}$ by products ${\prod_j \tilde X_j}$. Equivalently, this can be understood as a relabelling of the logical qubits in terms of a basis change to the readout basis. Note that this can change some properties of the problem graph and for example result in different parameters $\bm{\beta}$ and $\bm{\gamma}$. For some problems (for example problems with known parameter concentration~\cite{Brandao2018, Streif2020, Akshay2021}) it can thus be beneficial to instead simply add the missing data qubits to the code.

Regardless of the system size, the depth for measurement-based parity QAOA is composed from 2 measurement steps, 4 CNOT steps, and 4 steps containing only single-qubit operations, as given in Table~\ref{tab:measurement_qaoa_resources}.

\begin{table}[t]
    \centering
    \setlength{\tabcolsep}{0.5em} 
    {\renewcommand{\arraystretch}{1.2}
    \normalsize
    \begin{tabular}{c|c|c|c}
     \multirow{2}{*}{Circuit constituent} & \multicolumn{3}{c}{Required circuit depth}\\
 & Measure & CNOT & Single qubit \tabularnewline \hline
         Encoding & $1$      & $4$  & $1$ \\
         $U_P$    & $0$      & $0$  & $1$ \\
         Decoding & $1$      & $0$  & $1$ \\
         $U_X$    & $0$      & $0$  & $1$ \\\hline
         \textbf{total} & \textbf{2} & \textbf{4} & \textbf{4}
    \end{tabular}
    }
    \caption{Depth required for the constituents of a single QAOA layer. Note that duration of single qubit gates is negligible compare to the duration of measurements or 2-qubit gates.}
    \label{tab:measurement_qaoa_resources}
\end{table}

\subsection{Quantum Fourier Transform with linear qubit overhead}
\begin{figure}[t]
    \centering
    \includegraphics[width=\columnwidth]{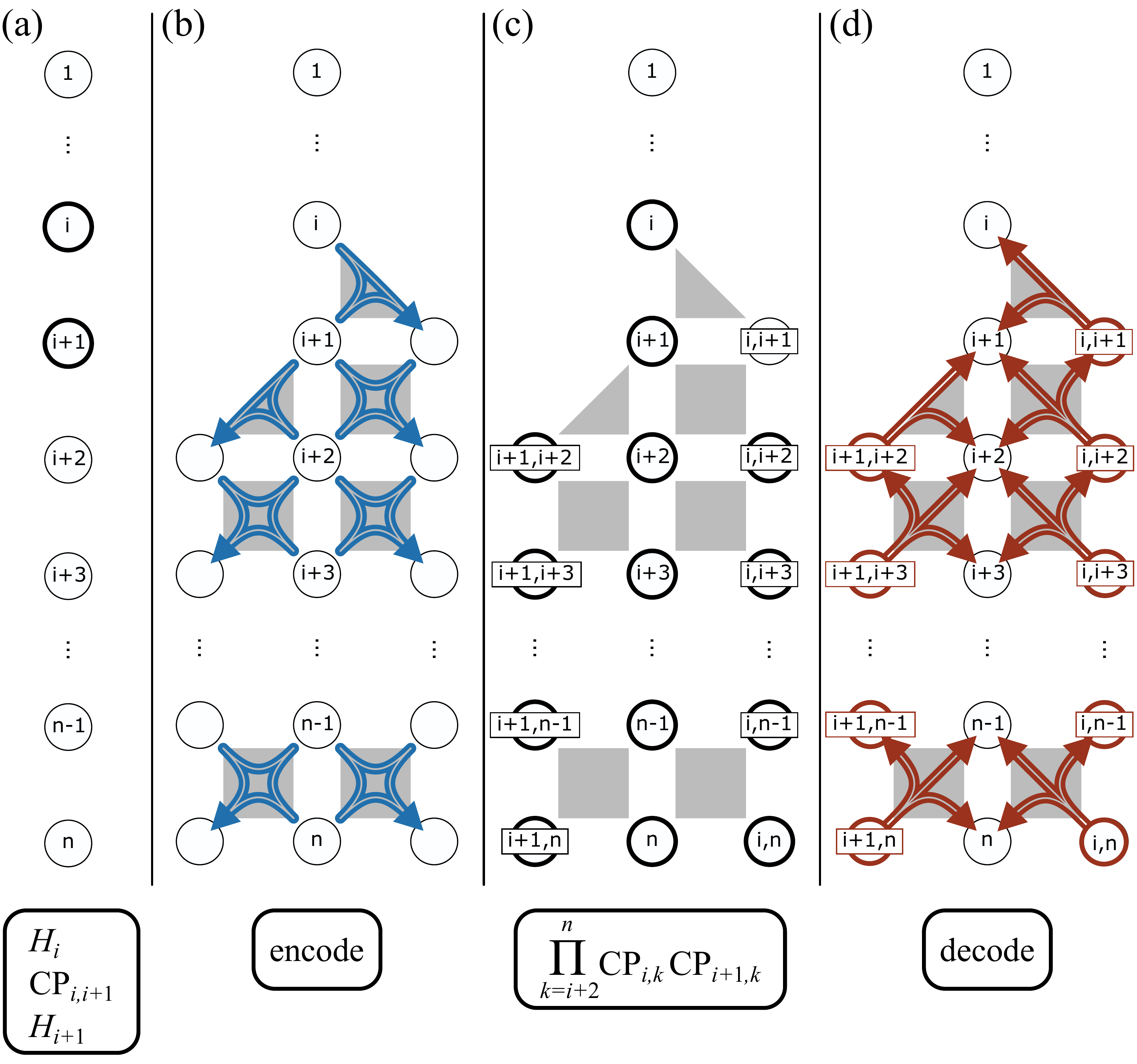}
    \caption{Basic block of the QFT with measurement-based encoding and decoding of parity qubits. The first part (a) consists of two Hadamard gates and a CP gate on the unencoded physical qubits. In the second step (b), parity qubits are added to the code in two lines around the data qubits to enable interactions of logical qubits $i$ and ${i+1}$ with all other logical qubits ${k>i+1}$. The corresponding CP gates are performed on the encoded state (c), before decoding all parity qubits again (d). This basic block is repeated $\lceil (n-2)/2 \rceil$ times. The last two logical gates of the QFT algorithm can be performed as physical gates on the unencoded qubits. Qubits that are subject to a quantum operation are highlighted with a bold outline.}
    \label{fig:QFT}
\end{figure}
In Ref.~\cite{Fellner2022applications} it has been shown how the quantum Fourier transform (QFT) on $n$ qubits can be implemented in the parity architecture using ${n(n+1)/2}$ qubits and a circuit depth of ${8n-9}$. Here we demonstrate a cost reduction by employing our new approach to dynamically encode and decode parity qubits tailored to specific parts of the QFT circuit, requiring a physical qubit layout corresponding to a square lattice of width 3 and length $n$ (i.e.,\ a linear amount of qubits, see Fig.~\ref{fig:QFT}) with an ancilla qubit in the center of every square. The middle line of qubits represents the logical qubits (as data qubits of the parity code), while the two outer lines are used to add parity qubits to implement the necessary interactions. The QFT requires implementing Hadamard gates ($H$) and controlled phase (CP) gates defined as ${\text{CP}(\phi) = {\rm diag}(1, 1, 1, e^{i\phi})}$ and is given by
\begin{equation}
    U_\text{QFT} = \prod_{i=1}^n \left[H_{(i)}\prod_{j=i+1}^n \text{CP}_{i, j}(2^{i-j}\pi)\right]. 
\end{equation}
We perform all Hadamard gates and the CP gates between neighboring qubits as physical gates on the data qubits, all remaining CP gates are implemented in an encoded state using physical single-body operations on the respective parity and data qubits (see Ref.~\cite{Fellner2022universal} for the exact implementation).
We apply the following block of instructions ${\lceil (n-2)/2 \rceil}$ times, starting at ${i=1}$:

\begin{itemize}
    \item [(a)] Perform the gates $H_{(i)}$, $ \text{CP}_{(i),(i+1)}(\pi/2)$ and $ H_{(i+1)}$  as physical gates on the corresponding data qubits.
    \item [(b)] Encode parity qubits ${(i,k)}$ and ${(i+1, l)}$ for all $k, l$ with ${i<k\leq n}$ and ${i+1<l\leq n}$ using measurement of plaquette constraints as shown in Fig.~\ref{fig:QFT}b.
    \item [(c)] Perform the logical gates ${\Tilde{\text{CP}}_{i,k}(2^{i-k} \pi)}$ and ${\Tilde{\text{CP}}_{i+1,k}(2^{i+1-k} \pi)}$ for all $k$ with ${i+1<k\leq n}$.
    \item [(d)] Decode by measuring all parity qubits.
    \item [(e)] Reassign ${i \gets i+2}$.
\end{itemize}
For even $n$, the last two gates of the QFT circuit are applied physically on the data qubits in the end.
Each block can be implemented in the depth of one $\text{CP}$ gate and four CNOT gates, or equivalently, six CNOT gates. In addition, each block requires two steps of simultaneous measurements and classical post-processing. 
For even $n$, this results in total depth of ${3n-4}$ CNOT steps and ${n-2}$ measurement steps (we omit steps with only single-qubit gates). For odd $n$, the circuit consists of ${3n-3}$ CNOT steps and ${n-1}$ measurement steps. Given the depth of ${10n-13}$ for the conventional implementation~\cite{Holmes2020} on a square lattice, measurement-based encoding and decoding can significantly speed up the QFT whenever fast projective measurements are available.

\subsection{Preparation of graph states}
As suggested in Ref.~\cite{Fellner2022applications}, the parity code can be used to create arbitrary graph states~\cite{Raussendorf2001, Raussendorf2003, Hein2006} of $n$ qubits in a circuit depth linear in $n$. With the measurement-based approach to encoding and decoding, this depth reduces to a constant: State preparation with single-qubit gates, encoding in four steps of CNOT gates and one measurement and corrections step, logical CZ gates in one step of single-qubit gates, and decoding with one measurement and correction step. To compare the resources to those of conventional approaches of graph state generation~\cite{Hoyer2006}, one must take into account that the number of required constraint measurements in our approach scales linearly with the connectivity of the desired graph state. Although  the initial encoded state in the parity code is not a cluster or graph state, the encoding procedure has some similarities with cluster state preparation~\cite{Raussendorf2001} and the exact relationship between this implementation and the generation of graph states via cluster states remains to be investigated.

\subsection{Arbitrary quantum algorithms}
The measurement-based protocol presented in this work also offers the potential for depth reductions of general quantum circuits. It is in particular promising for quantum circuits which can be grouped into blocks whose unitary is diagonal in either the $X$-, $Y$- or $Z$-basis (or any basis which transforms to the computational basis only by single-qubit operations). For example, every operator of the form
\begin{equation}
    U = \prod_k e^{i\alpha_k \prod_j \sigma_j},
\end{equation}
containing only one kind of Pauli operators ${\sigma_j \in \{X_j, Y_j, Z_j\}}$ but arbitrarily high orders of interaction, can be implemented in constant circuit depth (given that enough qubits are available to encode the required connectivity).The number of additional qubits always scales with the number of interactions to be implemented. In the best case, every interaction term requires two additional qubits (one measurement ancilla and one parity qubit). In some cases, however, additional ancillary parity qubits may need to be added in order to obtain a layout with local constraints. For blocks with $Y$ or $X$ interactions, the encoding needs to be performed in the corresponding basis, or equivalently, the appropriate basis change needs to be applied before and after the block.
This makes the parity code interesting in the context of synthesizing arbitrary quantum gates and algorithms from Ising-type interactions and single-body gates, as for example in Ref.~\cite{Bassler2022}.

Another field of application is the implementation of CZ circuits, which have constant depth in our approach. That is a significant improvement compared to the lowest known upper bound for implementations without the parity code, which is logarithmic in the number of qubits~\cite{Maslov2022}. However, note that the gain in circuit depth in our approach is related to an increase in circuit size, while the procedure in Ref.~\cite{Maslov2022} does not require any ancilla qubits.

\section{Conclusion}\label{sec:conclusion}
In this work, we have presented a novel way to encode and decode information in the parity encoding using a measurement-based technique, similar to the concept of lattice surgery in the surface code~\cite{Horsman2012}. The method can be used to arbitrarily change the shape of the code while keeping information on the logical qubits untouched. This allows for a flexible use of the parity code for a variety of quantum algorithms by dynamically adapting the code to the algorithmic requirements in a constant depth independent of the system size. In particular, we describe a constant-depth QAOA protocol in the parity architecture which does not require any energy penalization of parity constraints. A single QAOA layer for optimization problems on graphs and hypergraphs of arbitrary connectivity now runs in 10 time steps (containing single-qubit gates, two-qubit gates and measurements). This opens up a yet unexplored possibility to efficiently solve combinatorial optimization problems on qubit platforms with limited connectivity but efficient qubit measurement. The depth reduction of the quantum Fourier transform to at most ${4n-4}$ steps (measurements and CNOT gates) further suggests the potential for more, unexplored advantages of the parity code in quantum computing. 
Variational quantum algorithms are particularly interesting candidates for this, a deeper investigation into more general variational algorithms will be the scope of future work. While the parity code is not a full quantum error correction code, and deformations to small layouts further reduce the robustness to noise, we stress that the possibility to partially correct bit-flip errors via stabilizer measurements may still be harnessed in combinations of this setup with other error correction codes to reach full fault-tolerance.

\section{Acknowledgements}
Work at the University of Innsbruck was supported by the Austrian Science Fund (FWF) through a START grant under Project No. Y1067-N27 and the Special Research Programme (SFB) BeyondC Project No. F7108-N38. This project was funded within the QuantERA II Programme that has received funding from the European Union's Horizon 2020 research and innovation programme under Grant Agreement No. 101017733. The work was further supported by the Austrian Research Promotion Agency under Grant (FFG Project No. 892576, Basisprogramm).

\IEEEtriggeratref{39}

\end{document}